\documentstyle[12pt]{article}

\begin{document}
\renewcommand{\thefootnote}{\fnsymbol{footnote}}
\begin{titlepage}
\begin{flushright}
OU-HET 429 \\
hep-th/0301233
\end{flushright}

\vspace{10mm}
\begin{center}
{\Large\bf Parent Actions, Dualities and New Weyl-invariant Actions
of Bosonic $p$-branes}
\vspace{25mm}

{\large
Yan-Gang Miao$^{a,b,}$\footnote{e-mail address:
 miao@het.phys.sci.osaka-u.ac.jp}
and
Nobuyoshi Ohta$^{a,}$\footnote{e-mail address: ohta@phys.sci.osaka-u.ac.jp}}\\
\vspace{10mm}
${}^a$ {\em Department of Physics, Osaka University,
Toyonaka, Osaka 560-0043, Japan} \\
${}^b$ {\em Department of Physics, Xiamen University, Xiamen 361005,}\\
{\em People's Republic of China}
\end{center}
\vspace{15mm}
\centerline{{\bf{Abstract}}}
\vspace{5mm}

By using the systematic approach of parent action method, we derive
one Weyl-noninvariant and two Weyl-invariant actions of bosonic $p$-branes
($p\geq 2$) starting from the Nambu-Goto action, and establish the duality
symmetries in this set of four actions. Moreover, we discover a new bosonic
$p$-brane action (including the string theory) and deduce two corresponding
Weyl-invariant formulations by proposing a new special parent action.
We find that the same duality symmetries as those mentioned above exist
in this new set of actions.
The new $p$-brane actions are also briefly analyzed.

\end{titlepage}
\newpage
\renewcommand{\thefootnote}{\arabic{footnote}}
\setcounter{footnote}{0}
\setcounter{page}{2}

\section{Introduction}

The parent or master action approach was proposed by Deser and Jackiw~\cite{s1}
nearly two decades ago. The purpose was to establish, at the level of the
Lagrangian instead of equations of motion, the equivalence or so-called
duality between the Abelian self-dual and Maxwell-Chern-Simons models in
(2+1)-dimensional spacetime. The approach was in fact originated from the
Legendre transformation and recently has been applied and developed in
diverse directions. For instance, one direct development~\cite{s2} is that
it has been generalized to establish the duality between the non-Abelian
self-dual and Yang-Mills-Chern-Simons models. Another interesting
application~\cite{s3} is that the self-duality of various chiral
$p$-form actions has been established with the modification~\cite{s4} of
the approach with one more auxiliary field to preserve the manifest
Lorentz invariance of chiral $p$-forms.

The main idea of the parent action approach~\cite{s5} contains two steps:
(1) to introduce auxiliary fields and then construct a parent or master
action based on a known action, and (2) to make variation of the parent
action with respect to each auxiliary field, solve one auxiliary field
in terms of other fields and then substitute the solution into the parent
action. Through making variations with respect to different auxiliary fields,
we can obtain different forms of the actions. The actions are, of
course, equivalent classically, and the relation between them is usually
referred to as duality. If the resulting actions are the same,
the relation is called self-duality.

The bosonic $p$-branes~\cite{s6} are extended objects that are embedded in
a higher dimensional spacetime, and their dynamics is governed in general by
the following four different kinds of actions. The first is the
Nambu-Goto action ($S_{NG}$) that is proportional to the $(p+1)$-dimensional
worldvolume by Nambu~\cite{s8} and Goto~\cite{s9} for a string ($p=1$).
The second action is the one with an auxiliary worldvolume metric and
a cosmological term ($S_P$, where the subscript means $p$-branes), whose
formulation of a membrane ($p=2$) was proposed by Howe and Tucker~\cite{s10}
in the construction of a locally supersymmetric invariant model of
a spinning membrane. The two actions, $S_{NG}$ and $S_P$, are classically
equivalent to each other. Instead of the nonpolynomial
form of the former, the latter is quadratic in the derivatives of
spacetime coordinates at the price of introducing the auxiliary
worldvolume metric. Quite noticeable is that $S_P$ is not Weyl-invariant
for the general cases, $p \geq 2$, but Weyl-invariant for the string theory,
{\em i.e.}, $p=1$. This string action was first given by Brink, Di Vecchia and
Howe and Deser and Zumino~\cite{s11,s12}.
The other two $p$-brane actions possess a Weyl symmetry as demonstrated in
the following. Because of the important role played by the Weyl symmetry
in the covariant quantization of strings, the construction of $p$-brane
actions with the Weyl invariance was then motivated. One recent proposal
was to introduce an auxiliary scalar field in both the worldvolume and
spacetime in the Weyl-noninvariant formulation $S_P$, and to propose a
Weyl-invariant action of $p$-branes ($S^{\rm I}_{W}$, called the first
Weyl-invariant action in our paper) and a Weyl-invariant action of
$Dp$-branes as well~\cite{s15,s16,s14}. Of course, this Weyl-invariant
$p$-brane action is classically equivalent to that constructed long time
ago~\cite{s17,s6} by the simple expedient of raising the action of Brink et al.
for the string theory to the power $(p+1)/2$. The old Weyl-invariant action
($S^{\rm II}_{W}$, called the second Weyl-invariant action) is nonpolynomial
as the Nambu-Goto formulation, but contains no other auxiliary fields
besides the auxiliary worldvolume metric.

As we know, the relation revealed at present among the four different
$p$-brane actions is merely classical equivalence,\footnote{The duality was
discussed~\cite{s18} in the context of first- and second-order actions.}
that is, the same equations of motion as those from the original action
$S_{NG}$ can be derived from $S_P$, $S^{\rm I}_{W}$, and $S^{\rm II}_{W}$.
We may classify the four actions by polynomiality and by the
Weyl symmetry. For the polynomiality, $S_P$ and $S^{\rm I}_{W}$
belong to the class with quadratic derivatives of
spacetime coordinates while $S_{NG}$ and $S^{\rm II}_{W}$
constitute the class with nonpolynomiality; for the Weyl symmetry\footnote{We
do not consider the Weyl invariance of the Nambu-Goto action because it does
not contain the worldvolume metric.}
$S_P$ belongs to the class without the Weyl
invariance while $S^{\rm I}_{W}$ and $S^{\rm II}_{W}$ constitute the class
with such an invariance.

Because $p$-branes (including strings) can be described by different kinds
of actions it has been a noticeable topic to derive from one formulation
(usually the Nambu-Goto) the others systematically and to establish relations
among them, such as duality. To this end a first-order action method was
proposed~\cite{s18}, and the string action of Brink et al. with the Weyl
invariance was derived from the Nambu-Goto action of strings\footnote{For
$p$-brane ($p \geq 2$) cases a Weyl-invariant action was
constructed~\cite{s18} from the Nambu-Goto through directly replacing the
induced metric by a
Weyl-invariant combination of the worldvolume metric and the induced metric
itself. We do not include this action in our paper.} and the duality between
the two formulations was established. The main idea of the method coincides
with that of the parent action approach mentioned above, in both of which it
is inevitable to introduce Lagrange multipliers and other auxiliary fields
and then to make variations with respect to them. This method was also
applied to $Dp$-branes~\cite{s18,s19}. Recently another attempt, {\em i.e.},
a mechanism was suggested~\cite{s16} to derive the Weyl-invariant actions
of $p$-branes and $Dp$-branes. Here it should be pointed out that this
mechanism is in fact an interesting application of the parent action
approach as will be seen clearly in the following contents.

In this paper,
we start with the Nambu-Goto action of bosonic $p$-branes and derive
other $p$-brane actions by the systematic approach of parent action method.
Simultaneously we also establish the duality symmetries among the four
different $p$-brane actions. Incidentally, dual actions of
$Dp$-branes were discussed in refs.~\cite{s20,s21}, but the discussions were
limited to duality with respect to gauge fields. We note that this systematic
method is powerful {\em not only} in the derivation of known actions
{\em but also} in the discovery of new actions. In fact, by this method
we propose a special parent action and find out a new $p$-brane action
including the string theory. Furthermore, we obtain two corresponding
Weyl-invariant actions and establish the same duality symmetries among
the Nambu-Goto and three new actions as that among the four known actions.

Though the Nambu-Goto action of strings has geometrical meaning and is
intuitively easy to understand, it is not very convenient for covariant
quantization of the string. On the other hand, the string action of Brink
et al. does not have geometrical
meaning but is quite useful in the covariant quantization of strings.
Thus it is useful to make various formulations of the actions. It remains
to be seen if our new actions shed any insight in this direction.

The arrangement of this paper is as follows. In the next section, we first
write down~\cite{s16} the parent action of the Nambu-Goto formulation and
derive $S_P$ as the dual action of $S_{NG}$.
Secondly we construct the parent action of $S_P$ and
obtain $S_{NG}$ as the dual action of $S_P$.
Thereby we establish the duality between $S_{NG}$ and $S_P$
with respect to the induced metric and worldvolume metric,
respectively. In sect.~3, our discussions turn to the Weyl-invariant
$p$-brane actions. Introducing an auxiliary scalar field $\Phi$ in both
the worldvolume and spacetime and making the transformation $g_{ij}
\to {\Phi}g_{ij}$, where $g_{ij}$ is the worldvolume metric, in the
parent action of the Nambu-Goto formulation, we can write down a different
parent action~\cite{s16}. The variation of this parent action leads to
the first Weyl-invariant action as the dual action of the Nambu-Goto.
Moreover, constructing the parent action of the first Weyl-invariant
formulation, we derive the second Weyl-invariant action as the dual action
of the first Weyl-invariant one. As a result, the Nambu-Goto and first
Weyl-invariant actions are dual to each other with respect to the induced
metric, and the first and second Weyl-invariant actions are dual to each
other with respect to the scalar field. In sect.~4, we discuss the dual
actions of the first and second Weyl-invariant formulations with respect to
the worldvolume metric, and find that both of them are dual to $S_{NG}$,
but not to $S_P$ as is naively thought. In sect.~5, a special
parent action of the Nambu-Goto formulation is proposed by defining the
inverse of the induced metric. Completely following the procedure
described in the preceding three sections, we deduce a new action
that is dual to the Nambu-Goto and its two corresponding Weyl-invariant
ones, and establish the same duality symmetries among the Nambu-Goto and
three new actions as those among the four known actions. Finally
a conclusion is made in sect.~6.

The notation we use throughout this paper is as follows:
\begin{equation}
{\eta}_{{\mu}{\nu}}={\rm diag}(-1,1,\cdots,1),
\end{equation}
stands for the flat metric of the $D$-dimensional Minkowski spacetime.
Greek indices (${\mu},{\nu},{\sigma},\cdots$) run over $0,1,\cdots,D-1$.
\begin{equation}
h_{ij}\equiv \frac{{\partial}X^{\mu}}{{\partial}{\xi}^{i}}
\frac{{\partial}X^{\nu}}{{\partial}{\xi}^{j}}{\eta}_{{\mu}{\nu}},
\end{equation}
is defined as the induced metric in the $(p+1)$-dimensional worldvolume
spanned by $p+1$ arbitrary parameters ${\xi}^i$. Latin indices
($i,j,k,\cdots$) take the values $0,1,\cdots,p$. Coordinates
$X^{\mu}({\xi}^i)$ are the $D$-component Lorentz vector field in the
spacetime and $D$ scalar fields in the worldvolume as well.
Finally $g_{ij}$ means the auxiliary worldvolume metric and
$g^{ij}$ its inverse.

\section{Duality between $S_{NG}$ and $S_P$}

We begin with the Nambu-Goto action of bosonic $p$-branes
\begin{equation}
S_{NG}=-T_{p}\int d^{p+1}{\xi}\sqrt{-h},
\end{equation}
where $h\equiv {\rm det}(h_{ij})$ and $T_p$ is the $p$-brane tension.
According to the parent action approach~\cite{s1,s5}, we introduce
the worldvolume second-rank tensor fields ${\Lambda}^{ij}$ and $g_{ij}$,
and write down the parent action of the Nambu-Goto formulation
\begin{equation}
S^{{\rm I}}_{parent}=-T_{p}\int d^{p+1}{\xi}\left[\sqrt{-g}+{\Lambda}^{ij}
\left(g_{ij}-h_{ij}\right)\right],
\end{equation}
where $g\equiv {\rm det}(g_{ij})$. We will see later that $g_{ij}$ is
just the auxiliary worldvolume metric, but at present it is treated as
an independent auxiliary field. Note that eq.~(4)\footnote{The string case
of this equation appeared in ref.~\cite{s18}.} was obtained in
ref.~\cite{s16} by the consideration that coincides with the parent action
approach.

Now varying eq.~(4) with respect to ${\Lambda}^{ij}$ gives the relation
$g_{ij}=h_{ij}$, together with which eq.~(4) turns back to the Nambu-Goto
action~(3). This shows the classical equivalence between the parent and
Nambu-Goto actions. However, varying eq.~(4) with respect to $g_{ij}$
leads to the expression of ${\Lambda}^{ij}$ in terms of $g_{ij}$:
\begin{equation}
{\Lambda}^{ij}=-\frac{1}{2}\sqrt{-g}g^{ij},
\end{equation}
where $g^{ij}$ is the inverse of $g_{ij}$. Substituting eq.~(5) into eq.~(4),
we obtain the dual version of the Nambu-Goto action
\begin{equation}
S_{P}=-\frac{T_p}{2}\int d^{p+1}{\xi}\sqrt{-g}\left[g^{ij}h_{ij}-(p-1)\right].
\label{npol}
\end{equation}
This is the $p$-brane action with the auxiliary field $g_{ij}$ that now
plays the role of the worldvolume metric and with the cosmological term
for $p\geq 2$. For the string theory ($p=1$), it reduces to the
Brink-Di Vecchia-Howe-Deser-Zumino action as we noted in the previous
section. In the above, we have shown that
$S_P$ is dual to $S_{NG}$ with respect to the induced metric $h_{ij}$.
We note, however, that eq.~(\ref{npol}) is not Weyl-invariant for $p\geq 2$.

Parent actions are not unique. Starting from $S_P$,
we construct the parent action
\begin{equation}
S^{{\rm II}}_{parent}=-\frac{T_p}{2}\int d^{p+1}{\xi}\left[\sqrt{-G}
\left(G^{ij}h_{ij}-(p-1)\right)+{\Lambda}_{ij}\left(G^{ij}-g^{ij}\right)
\right],
\end{equation}
where ${\Lambda}_{ij}$ and $G^{ij}$ are two auxiliary second-rank tensor
fields introduced in the worldvolume, $G\equiv {\rm det}(G_{ij})$ and
$G_{ij}$ is the inverse of $G^{ij}$, and vice versa. We can verify that
$S_{NG}$ is dual to $S_P$ with respect
to the inverse of the worldvolume metric $g^{ij}$. The procedure of the
verification is as follows. First making the variation of eq.~(7) with
respect to ${\Lambda}_{ij}$ gives $G^{ij}=g^{ij}$, with which eq.~(7) reduces
to the $p$-brane action~(\ref{npol}). This means that the second parent action
is classically equivalent to $S_P$. Secondly, to make the
variation of eq.~(7) with respect to $G^{ij}$ leads to the relation of
${\Lambda}_{ij}$ expressed by $h_{ij}$ and $G_{ij}$:
\begin{equation}
{\Lambda}_{ij}=\frac{1}{2}\sqrt{-G}G_{ij}\left(G^{kl}h_{kl}-(p-1)\right)
-\sqrt{-G}h_{ij},
\end{equation}
which is independent of $g^{ij}$. After substituting eq.~(8) into eq.~(7)
and making the variation again with respect to $g^{ij}$ which is now dealt
with as a Lagrangian multiplier, we simply derive ${\Lambda}_{ij}=0$, that is,
\begin{equation}
\frac{1}{2}G_{ij}\left(G^{kl}h_{kl}-(p-1)\right)-h_{ij}=0.
\end{equation}
The solution for the $p\geq 2$ cases is
\begin{equation}
G_{ij}=h_{ij}.
\end{equation}
Using eqs.~(8)-(10), we then obtain from eq.~(7) the Nambu-Goto action.
As for the string case, {\em i.e.}, $p=1$, eq.~(9) is simplified to be
\begin{equation}
\frac{1}{2}G_{ij}G^{kl}h_{kl}-h_{ij}=0,
\end{equation}
whose solution is
\begin{equation}
G_{ij}=f_{1}({\xi}^{0},{\xi}^{1})h_{ij},
\end{equation}
where $f_{1}({\xi}^{0},{\xi}^{1})$ is an arbitrary scalar function of the
worldsheet parameters. This arbitrary function appears due to the Weyl
invariance existing in the action of Brink et al. for the string theory.
By using eqs.~(8), (11) and (12) and the condition $p=1$, we have from
eq.~(7) the Nambu-Goto action for strings as expected. Hence
the dual action deduced from the second parent action is exactly the
Nambu-Goto action as we have claimed.

The duality established above may be illustrated as in Figs.1 and 2
as is done in ref.~\cite{s5}.
\begin{figure}[htb]
\begin{minipage}{.45\linewidth}
\begin{center}
\setlength{\unitlength}{.7mm}
\begin{picture}(100,55)(10,5)
\put(46,60){$S^{{\rm I}}_{parent}$}
\put(46,57){\vector(-2,-3){23}}\put(54,57){\vector(2,-3){23}}
\put(22,40){${\delta {\Lambda}^{ij}}$}\put(68,40){${\delta g_{ij}}$}
\put(18,17){${S_{NG}}$}\put(78,17){${S_{P}}$}
\put(20,10){\vector(0,1){5}}\put(80,10){\vector(0,1){5}}
\put(20,10){\line(1,0){20}}\put(60,10){\line(1,0){20}}
\put(46,9){\bf{dual}}
\end{picture}
\caption{\small Schematic relation of the actions: The parent action
$S^{{\rm I}}_{parent}$ is classically equivalent to $S_{NG}$ and $S_{P}$,
and $S_{NG}$ and $S_{P}$ are dual to each other with respect to the
induced metric.}
\label{dual1-1}
\end{center}
\end{minipage}
\hspace{10mm}
\begin{minipage}{.45\linewidth}
\begin{center}
\setlength{\unitlength}{.7mm}
\begin{picture}(100,50)(15,5)
\put(46,60){$S^{{\rm II}}_{parent}$}
\put(46,57){\vector(-2,-3){23}}\put(54,57){\vector(2,-3){23}}
\put(22,40){${\delta {\Lambda}_{ij}}$}\put(68,40){${\delta G^{ij}}
+{\delta g^{ij}}$}
\put(18,17){${S_{P}}$}\put(78,17){${S_{NG}}$}
\put(20,10){\vector(0,1){5}}\put(80,10){\vector(0,1){5}}
\put(20,10){\line(1,0){20}}\put(60,10){\line(1,0){20}}
\put(46,9){\bf{dual}}
\end{picture}
\caption{\small Schematic relation of the actions:
The parent action $S^{{\rm II}}_{parent}$ shows that $S_{P}$ and $S_{NG}$
are dual to each other with respect to the worldvolume metric.}
\end{center}
\end{minipage}
\end{figure}

\section{Dualities between $S_{NG}$ and $S^{\rm I}_{W}$ and between
$S^{\rm I}_{W}$ and $S^{\rm II}_{W}$}

Introducing an auxiliary scalar field ${\Phi}({\xi}^i)$, and rescaling
the worldvolume metric $g_{ij}\to{\Phi}g_{ij}$ in the parent
action of the Nambu-Goto~(4), we write down the third parent action~\cite{s16}
\begin{equation}
S^{{\rm III}}_{parent}=-T_{p}\int d^{p+1}{\xi}\left[{\Phi}^{(p+1)/2}
\sqrt{-g}+{\Lambda}^{ij}\left({\Phi}g_{ij}-h_{ij}\right)\right],
\end{equation}
where ${\Phi}({\xi}^i)$ should be a scalar in both the spacetime and
worldvolume in order to keep eq.~(13) invariant under the Lorentz
transformation and reparametrization.

Varying eq.~(13) with respect to ${\Lambda}^{ij}$ brings about
$g_{ij}={\Phi}^{-1}h_{ij}$, which leads to nothing new but the classical
equivalence between the Nambu-Goto and third parent actions. However,
varying the equation with respect to $g_{ij}$, we get ${\Lambda}^{ij}$ as
\begin{equation}
{\Lambda}^{ij}=-\frac{1}{2}{\Phi}^{(p-1)/2}\sqrt{-g}g^{ij}.
\end{equation}
Substituting eq.~(14) back to eq.~(13), we derive the dual action
\begin{equation}
S^{{\rm I}}_{W}=-\frac{T_p}{2}\int d^{p+1}{\xi}\sqrt{-g}\left[{\Phi}^{(p-1)/2}
g^{ij}h_{ij}-(p-1){\Phi}^{(p+1)/2}\right],
\end{equation}
which was obtained in ref.~\cite{s16} but whose duality was not uncovered.
This action is interesting because it is invariant under the Weyl
transformation
\begin{eqnarray}
X^{\mu}({\xi}) & \longrightarrow & X^{\mu}({\xi}), \nonumber \\
g_{ij}({\xi}) & \longrightarrow & {\rm exp}\left({\omega}({\xi})\right)
g_{ij}({\xi}), \nonumber \\
{\Phi}({\xi}) & \longrightarrow & {\rm exp}\left(-{\omega}({\xi})\right)
{\Phi}({\xi}),
\label{wt}
\end{eqnarray}
where ${\omega}({\xi})$ is an arbitrary real function of worldvolume
parameters. We thus call it the first Weyl-invariant $p$-brane action in
this paper. For the string theory, {\em i.e.}, $p=1$, $S^{{\rm I}}_{W}$
turns back to the Weyl-invariant action of Brink et al. The duality between
$S_{NG}$ and $S^{{\rm I}}_{W}$ may be illustrated as in Fig.~3.
\begin{figure}[htb]
\begin{center}
\setlength{\unitlength}{.7mm}
\begin{picture}(100,50)(0,10)
\put(46,60){$S^{{\rm III}}_{parent}$}
\put(46,57){\vector(-2,-3){23}}\put(54,57){\vector(2,-3){23}}
\put(22,40){${\delta {\Lambda}^{ij}}$}\put(68,40){${\delta g_{ij}}$}
\put(18,17){${S_{NG}}$}\put(78,17){${S^{{\rm I}}_{W}}$}
\put(20,10){\vector(0,1){5}}\put(80,10){\vector(0,1){5}}
\put(20,10){\line(1,0){20}}\put(60,10){\line(1,0){20}}
\put(46,9){\bf{dual}}
\end{picture}
\caption{\small Schematic relation of the actions: The parent action
$S^{{\rm III}}_{parent}$ shows that $S_{NG}$ and $S^{{\rm I}}_{W}$ are
dual to each other with respect to the induced metric.}
\end{center}
\end{figure}
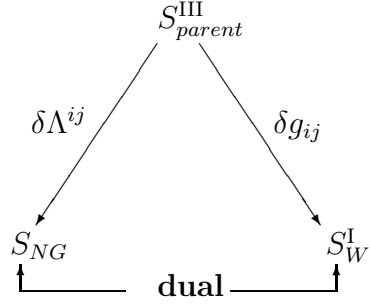

Next we introduce two auxiliary scalar fields $\Lambda$ and $\Psi$, and
further construct the fourth parent action which corresponds to
$S^{{\rm I}}_{W}$
\begin{equation}
S^{{\rm IV}}_{parent}=-\frac{T_p}{2}\int d^{p+1}{\xi}\left[\sqrt{-g}
\left({\Psi}^{(p-1)/2}g^{ij}h_{ij}-(p-1){\Psi}^{(p+1)/2}\right)
+{\Lambda}\left({\Psi}-{\Phi}\right)\right].
\end{equation}
It is trivial to make the variation of eq.~(17) with respect to $\Lambda$,
which just leads to the classical equivalence between $S^{{\rm IV}}_{parent}$
and $S^{{\rm I}}_{W}$. However, it is nontrivial to vary eq.~(17) with
respect to $\Psi$, from which we deduce the relation
\begin{equation}
{\Lambda}=-\frac{p-1}{2}{\Psi}^{(p-3)/2}\sqrt{-g}\left(g^{ij}h_{ij}
-(p+1){\Psi}\right),
\end{equation}
which is independent of $\Phi$. Substituting eq.~(18) into eq.~(17) and
varying the parent action again with respect to $\Phi$ treated as
a Lagrangian multiplier now, we obtain ${\Lambda}=0$, that is,
\begin{equation}
{\Psi}=\frac{1}{p+1}g^{ij}h_{ij},
\end{equation}
for the $p\geq 2$ cases. Using eqs.~(18) and (19), we then derive
from $S^{{\rm IV}}_{parent}$ the dual action
\begin{equation}
S^{{\rm II}}_{W}=-T_{p}\int d^{p+1}{\xi}\sqrt{-g}\left(\frac{1}{p+1}g^{ij}
h_{ij}\right)^{(p+1)/2}.
\end{equation}
This $p$-brane action~\cite{s17,s6} also preserves invariance under the Weyl
transformation~(16), and is called here the second Weyl-invariant
action that does not contain auxiliary scalar fields. For $p=1$,
eq.~(17) simply reduces to the Weyl-invariant action of Brink et al. for
the string theory. Therefore, the duality between $S^{{\rm I}}_{W}$
and $S^{{\rm II}}_{W}$ is established and may be illustrated as in Fig.~4.
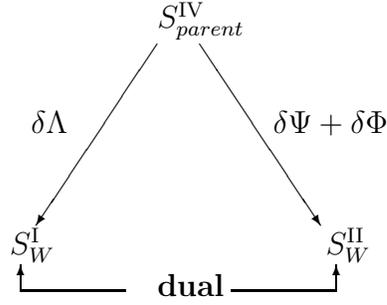
\begin{figure}[htb]
\begin{center}
\setlength{\unitlength}{.7mm}
\begin{picture}(100,55)(0,5)
\put(46,60){$S^{{\rm IV}}_{parent}$}
\put(46,57){\vector(-2,-3){23}}\put(54,57){\vector(2,-3){23}}
\put(22,40){${\delta {\Lambda}}$}\put(68,40){${\delta {\Psi}}+{\delta {\Phi}}$}
\put(18,17){${S^{{\rm I}}_{W}}$}\put(78,17){${S^{{\rm II}}_{W}}$}
\put(20,10){\vector(0,1){5}}\put(80,10){\vector(0,1){5}}
\put(20,10){\line(1,0){20}}\put(60,10){\line(1,0){20}}
\put(46,9){\bf{dual}}
\end{picture}
\caption{\small Schematic relation of the actions: The parent action
$S^{{\rm IV}}_{parent}$ shows that $S^{{\rm I}}_{W}$ and $S^{{\rm II}}_{W}$
are dual to each other with respect to the scalar field.}
\end{center}
\end{figure}

\section{Dualities between $S^{{\rm I}}_{W}$ and $S_{NG}$ and between $S^{{\rm II}}_{W}$ and $S_{NG}$}

In order to acquire a whole understanding of the dualities among $S_{NG}$,
$S_{P}$, $S^{{\rm I}}_{W}$, and $S^{{\rm II}}_{W}$, we have to exhaust all
parent actions. To this end, we construct the remaining parent actions
\begin{equation}
S^{{\rm V}}_{parent}=-\frac{T_p}{2}\int d^{p+1}{\xi}\left[\sqrt{-G}
\left({\Phi}^{(p-1)/2}G^{ij}h_{ij}-(p-1){\Phi}^{(p+1)/2}\right)
+{\Lambda}_{ij}\left(G^{ij}-g^{ij}\right)\right],
\end{equation}
and
\begin{equation}
S^{{\rm VI}}_{parent}=-T_{p}\int d^{p+1}{\xi}\left[\sqrt{-G}
\left(\frac{1}{p+1}G^{ij}h_{ij}\right)^{(p+1)/2}+{\Lambda}_{ij}\left(G^{ij}
-g^{ij}\right)\right].
\end{equation}
They correspond to $S^{{\rm I}}_{W}$ and $S^{{\rm II}}_{W}$, respectively,
which can be easily seen by varying them with respect to ${\Lambda}_{ij}$.
On the other hand, varying with respect to $G^{ij}$, we calculate
${\Lambda}_{ij}$ from eq.~(21):
\begin{equation}
{\Lambda}_{ij}=\frac{1}{2}\sqrt{-G}G_{ij}\left({\Phi}^{(p-1)/2}G^{kl}h_{kl}
-(p-1){\Phi}^{(p+1)/2}\right)-\sqrt{-G}{\Phi}^{(p-1)/2}h_{ij},
\end{equation}
and from eq.~(22):
\begin{equation}
{\Lambda}_{ij}=\frac {1}{2}\sqrt{-G}\left(\frac{1}{p+1}G_{ij}G^{kl}h_{kl}
-h_{ij}\right)\left(\frac{1}{p+1}G^{mn}h_{mn}\right)^{(p-1)/2}.
\end{equation}
It is important to note that both eq.~(23) and eq.~(24) are independent of
$g^{ij}$. Upon substituting eq.~(23) and eq.~(24) into $S^{{\rm V}}_{parent}$
and $S^{{\rm VI}}_{parent}$, respectively, and varying the parent actions
again with respect to $g^{ij}$ dealt with as Lagrangian multipliers,
we simply have ${\Lambda}_{ij}=0$ because of the independence of the
worldvolume metric. Therefore, eqs.~(23) and (24) are simplified to be
\begin{equation}
\frac{1}{2}G_{ij}\left(G^{kl}h_{kl}-(p-1){\Phi}\right)-h_{ij}=0,
\end{equation}
and
\begin{equation}
\frac{1}{p+1}G_{ij}G^{kl}h_{kl}-h_{ij}=0,
\end{equation}
respectively. Now let us solve eqs.~(25) and (26). Multiplying $G^{ij}$
on both sides of eq.~(25) and considering $p\geq 2$, we have
\begin{equation}
{\Phi}=\frac{1}{p+1}G^{ij}h_{ij}.
\end{equation}
By using this relation, we find that eq.~(25) reduces to eq.~(26).
Thus their solutions are
\begin{equation}
G_{ij}=f_{2}(\xi)h_{ij},
\end{equation}
where $f_{2}(\xi)$ is an arbitrary scalar function of the worldvolume
parameters ${\xi}^i$, and it exists because $S^{{\rm I}}_{W}$ and
$S^{{\rm II}}_{W}$ have the Weyl invariance as explained before.
Substituting eqs.~(23), (25), (27) and (28) into $S^{{\rm V}}_{parent}$,
and eqs.~(24), (26) and (28) into $S^{{\rm VI}}_{parent}$, we arrive at
the same dual action, the Nambu-Goto action~$S_{NG}$. As a result,
both $S^{{\rm I}}_{W}$ and $S^{{\rm II}}_{W}$ are dual to $S_{NG}$ with
respect to the worldvolume metric, and their dualities may be illustrated
as in Figs.~5 and 6.
\begin{figure}[htb]
\begin{minipage}{.45\linewidth}
\begin{center}
\setlength{\unitlength}{.7mm}
\begin{picture}(100,55)(10,10)
\put(46,60){$S^{{\rm V}}_{parent}$}
\put(46,57){\vector(-2,-3){23}}\put(54,57){\vector(2,-3){23}}
\put(22,40){${\delta {\Lambda}_{ij}}$}\put(68,40){${\delta G^{ij}}
+{\delta g^{ij}}$}
\put(18,17){${S^{{\rm I}}_{W}}$}\put(78,17){${S_{NG}}$}
\put(20,10){\vector(0,1){5}}\put(80,10){\vector(0,1){5}}
\put(20,10){\line(1,0){20}}\put(60,10){\line(1,0){20}}
\put(46,9){\bf{dual}}
\end{picture}
\caption{\small Schematic relation of the actions: The parent action
$S^{{\rm V}}_{parent}$ shows that $S^{{\rm I}}_{W}$ and $S_{NG}$ are dual
to each other with respect to the worldvolume metric.}
\end{center}
\end{minipage}
\hspace{10mm}
\begin{minipage}{.45\linewidth}
\begin{center}
\setlength{\unitlength}{.7mm}
\begin{picture}(100,55)(10,10)
\put(46,60){$S^{{\rm VI}}_{parent}$}
\put(46,57){\vector(-2,-3){23}}\put(54,57){\vector(2,-3){23}}
\put(22,40){${\delta {\Lambda}_{ij}}$}\put(68,40){${\delta G^{ij}}
+{\delta g^{ij}}$}
\put(18,17){${S^{{\rm II}}_{W}}$}\put(78,17){${S_{NG}}$}
\put(20,10){\vector(0,1){5}}\put(80,10){\vector(0,1){5}}
\put(20,10){\line(1,0){20}}\put(60,10){\line(1,0){20}}
\put(46,9){\bf{dual}}
\end{picture}
\caption{\small Schematic relation of the actions: The parent action
$S^{{\rm VI}}_{parent}$ shows that $S^{{\rm II}}_{W}$ and $S_{NG}$ are
dual to each other with respect to the worldvolume metric.}
\end{center}
\end{minipage}
\end{figure}

Starting from the Nambu-Goto action, we have derived the other three $p$-brane
actions $S_{P}$, $S^{{\rm I}}_{W}$, and $S^{{\rm II}}_{W}$ by using
our systematic parent action approach~\cite{s1,s5}. Simultaneously,
we have found all the parent actions that can be constructed from the
four actions, and have established their duality symmetries as shown in
Figs.~1-6. It is quite interesting that $S_{NG}$ and $S_{P}$, though dual
to each other, are not in an equal position even at the classical level
among the four dual actions, because $S^{{\rm I}}_{W}$ and
$S^{{\rm II}}_{W}$ are related to $S_{NG}$ with {\em direct} dualities
(see Figs.~3, 5 and 6) but not to $S_{P}$ with such dualities.
Thus the natural higher dimensional generalization of the action of Brink et
al. for the strings is the Weyl-invariant action (20) not only because it is
directly related to the Nambu-Goto action but also because the string action
of Brink et al. is Weyl invariant in contrast to the naive
generalization~(\ref{npol}).

We may precisely describe the well-known classical equivalence between
$S_{NG}$ and $S_{P}$ ``in the sense of equations of motion''.
On the other hand, we also notice the non-equality between $S_{NG}$ and
$S_{P}$ from the point of view of the number of Lagrangians.
If we set $\Phi$ to a constant $a$ in $S^{{\rm I}}_{W}$ in eq.~(15),
we obtain {\em an infinite number} (due to the arbitrary constant $a$)
of so-called $S_P$-type actions~\cite{s20}
\begin{equation}
S^{\prime}_{P}=-\frac{T^{\prime}_{p}}{2}\int d^{p+1}{\xi}\sqrt{-g}
\left[g^{ij}h_{ij}-(p-1)a\right],
\label{pol}
\end{equation}
where $T^{\prime}_{p}=a^{(p-1)/2}T_p$, and the $a=1$ case corresponds to $S_P$. They are classically equivalent to
the {\em unique} $S_{NG}$ after using the equations of motion.

\section{New bosonic $p$-brane actions and dualities}

In sect.~1, we mentioned that the parent action approach can also help
us to find new $p$-brane actions that are different from the four known ones.
Now we turn to the discussion in this aspect.

Let us begin with the Nambu-Goto action as the starting point of our
systematic method, and construct a new parent action as follows:
\begin{equation}
A^{{\rm I}}_{parent}=-T_{p}\int d^{p+1}{\xi}\left[\sqrt{-g}+{\Lambda}_{ij}
\left(g^{ij}-h^{ij}\right)\right].
\label{np1}
\end{equation}
At first sight it looks like $S^{{\rm I}}_{parent}$ in eq.~(4), just with
the exchange of lower and upper indices in the second term. We note that
if $h^{ij}$ is understood as $g^{ik}g^{jl}h_{kl}, A^{{\rm I}}_{parent}$ is
exactly the same as $S^{{\rm I}}_{parent}$ and nothing new can be deduced from
eq.~(\ref{np1}) but the $p$-brane action~(\ref{npol}). Actually $h^{ij}$ in
eq.~(\ref{np1}) is defined as the inverse of $h_{ij}$ and is independent of
$g_{ij}$. With this in mind, we follow the usual procedure described
in sect.~2 and derive a new $p$-brane action that is dual and equivalent
to $S_{NG}$:
\begin{equation}
A_{P}=-\frac{T_p}{2}\int d^{p+1}{\xi}\sqrt{-g}\left[-g_{ij}h^{ij}+(p+3)\right].
\label{np2}
\end{equation}
For the string theory ($p=1$), we can prove
\begin{equation}
g_{ij}h^{ij}=gh^{-1}g^{ij}h_{ij}, \hspace{10pt}(i,j=0,1),
\end{equation}
and then obtain a new action for strings
\begin{equation}
A_{P}(string)=-\frac{T_1}{2}\int d^{2}{\xi}\sqrt{-g}\left(-gh^{-1}
g^{ij}h_{ij}+4\right),\hspace{10pt}(i,j=0,1).
\end{equation}

The new $p$-brane action is Lorentz invariant and reparametrization
invariant as well. Under the reparametrization, the factor
$d^{p+1}{\xi}\sqrt{-g}$ remains unchanged and $g_{ij}$ and $h^{ij}$
transform as the worldvolume covariant and contravariant tensors, respectively,
\begin{eqnarray}
g_{ij}(\xi) =  \frac{{\partial}{\xi}^{{\prime}k}}{{\partial}{\xi}^i}
\frac{{\partial}{\xi}^{{\prime}l}}{{\partial}{\xi}^j}g^{\prime}_{kl}(
{\xi}^{\prime}), \nonumber \\
h^{ij}(\xi) =  \frac{{\partial}{\xi}^i}{{\partial}{\xi}^{{\prime}k}}
\frac{{\partial}{\xi}^j}{{\partial}{\xi}^{{\prime}l}}h^{{\prime}{kl}}(
{\xi}^{\prime}),
\end{eqnarray}
which keeps $g_{ij}h^{ij}$ invariant. Unfortunately, the action~(\ref{np2})
does not possess the Weyl invariance under (\ref{wt}) {\em even} for the
string theory. However, this invariance can be recovered by
introducing an auxiliary scalar field by the procedure done for $S_{P}$
in sect.~3, which will be shown below. On the other hand, $A_{P}$ and
$A_{P}(string)$ are nonpolynomial in the derivatives of spacetime
coordinates because of our special definition of $h^{ij}$. But this brings
in nothing strange as $S_{NG}$ possesses such a nonpolynomiality intrinsically.

For the new $p$-brane action, we can write down its corresponding parent action
\begin{equation}
A^{{\rm II}}_{parent}=-\frac{T_p}{2}\int d^{p+1}{\xi}\left[\sqrt{-G}
\left(-G_{ij}h^{ij}+(p+3)\right)+{\Lambda}^{ij}\left(G_{ij}-g_{ij}
\right)\right].
\end{equation}
Following the similar treatment as done in sect.~2 we arrive at $S_{NG}$
as the dual action. The duality between $S_{NG}$ and $A_{P}$ is thus
established and illustrated in Figs.~7 and 8.
\begin{figure}[htb]
\begin{minipage}{.45\linewidth}
\begin{center}
\setlength{\unitlength}{.7mm}
\begin{picture}(100,60)(10,10)
\put(46,60){$A^{{\rm I}}_{parent}$}
\put(46,57){\vector(-2,-3){23}}\put(54,57){\vector(2,-3){23}}
\put(22,40){${\delta {\Lambda}_{ij}}$}\put(68,40){${\delta g^{ij}}$}
\put(18,17){${S_{NG}}$}\put(78,17){${A_{P}}$}
\put(20,10){\vector(0,1){5}}\put(80,10){\vector(0,1){5}}
\put(20,10){\line(1,0){20}}\put(60,10){\line(1,0){20}}
\put(46,9){\bf{dual}}
\end{picture}
\caption{\small Schematic relation of the actions:
The parent action $A^{{\rm I}}_{parent}$ shows that $S_{NG}$ and $A_{P}$
are dual to each other with respect to the inverse of the induced metric.}
\end{center}
\end{minipage}
\hspace{10mm}
\begin{minipage}{.45\linewidth}
\begin{center}
\setlength{\unitlength}{.7mm}
\begin{picture}(100,60)(10,10)
\put(46,60){$A^{{\rm II}}_{parent}$}
\put(46,57){\vector(-2,-3){23}}\put(54,57){\vector(2,-3){23}}
\put(22,40){${\delta {\Lambda}^{ij}}$}\put(68,40){${\delta G_{ij}}
+{\delta g_{ij}}$}
\put(18,17){${A_{P}}$}\put(78,17){${S_{NG}}$}
\put(20,10){\vector(0,1){5}}\put(80,10){\vector(0,1){5}}
\put(20,10){\line(1,0){20}}\put(60,10){\line(1,0){20}}
\put(46,9){\bf{dual}}
\end{picture}
\caption{\small Schematic relation of the actions:
The parent action $A^{{\rm II}}_{parent}$ shows that $A_{P}$ and $S_{NG}$
are dual to each other with respect to the worldvolume metric.}
\end{center}
\end{minipage}
\end{figure}

Now let us restore the Weyl invariance of the new $p$-brane action. As was
done in sect.~3, introducing an auxiliary scalar field ${\Phi}({\xi}^i)$,
and rescaling the worldvolume metric as ${\Phi}g_{ij}$ in
$A^{{\rm I}}_{parent}$ in eq.~(\ref{np1}), we construct a new parent action
\begin{equation}
A^{{\rm III}}_{parent}=-T_{p}\int d^{p+1}{\xi}\left[{\Phi}^{(p+1)/2}
\sqrt{-g}+{\Lambda}_{ij}\left({\Phi}^{-1}g^{ij}-h^{ij}\right)\right].
\end{equation}
According to the steps of the parent action approach described in detail
previously, we deduce the dual action
\begin{equation}
A^{{\rm I}}_{W}=-\frac{T_p}{2}\int d^{p+1}{\xi}\sqrt{-g}
\left[-{\Phi}^{(p+3)/2}g_{ij}h^{ij}+(p+3){\Phi}^{(p+1)/2}\right].
\label{sw1}
\end{equation}
It is obviously invariant under the Weyl transformation~(16), and is
referred to as the first new Weyl-invariant action. As to the string case,
$A^{{\rm I}}_{W}$ simplifies to
\begin{equation}
A^{{\rm I}}_{W}(string)=-\frac{T_1}{2}\int d^{2}{\xi}\sqrt{-g}
\left(-{\Phi}^{2}gh^{-1}g^{ij}h_{ij}+4{\Phi}\right),\hspace{10pt}(i,j=0,1).
\end{equation}
It is possible to make its Hamiltonian analysis by the method in
ref.~\cite{s22} that is suitable to the nonpolynomial $S^{{\rm II}}_{W}$
in eq.~(20).

Next starting from $A^{{\rm I}}_{W}$, we construct a further parent action
\begin{equation}
A^{{\rm IV}}_{parent}=-\frac{T_p}{2}\int d^{p+1}{\xi}\left[\sqrt{-g}
\left(-{\Psi}^{(p+3)/2}g_{ij}h^{ij}+(p+3){\Psi}^{(p+1)/2}\right)
+{\Lambda}\left({\Psi}-{\Phi}\right)\right].
\end{equation}
By following the same procedure utilized to derive $S^{{\rm II}}_{W}$,
we obtain the action dual to $A^{{\rm I}}_{W}$:
\begin{equation}
A^{{\rm II}}_{W}=-T_{p}\int d^{p+1}{\xi}\sqrt{-g}\left(\frac{1}{p+1}g_{ij}
h^{ij}\right)^{-(p+1)/2},
\end{equation}
which is evidently invariant under the Weyl transformation~(\ref{wt}), and
is named the second new Weyl-invariant action. For the string theory,
$A^{{\rm II}}_{W}$ reduces to
\begin{equation}
A^{{\rm II}}_{W}(string)=2T_{1}\int d^{2}{\xi}\frac{h}{\sqrt{-g}g^{ij}h_{ij}},
\hspace{10pt}(i,j=0,1).
\end{equation}
Note that $A^{{\rm I}}_{W}(string)$ and $A^{{\rm II}}_{W}(string)$ have
different forms but, of course, dual and equivalent, while both
$S^{{\rm I}}_{W}$ and $S^{{\rm II}}_{W}$ lead to the string action of Brink
et al. The dualities between $S_{NG}$ and $A^{{\rm I}}_{W}$ and
between $A^{{\rm I}}_{W}$ and $A^{{\rm II}}_{W}$ are shown in Figs.~9 and
10, respectively.
\begin{figure}[htb]
\begin{minipage}{.45\linewidth}
\begin{center}
\setlength{\unitlength}{.7mm}
\begin{picture}(100,60)(10,10)
\put(46,60){$A^{{\rm III}}_{parent}$}
\put(46,57){\vector(-2,-3){23}}\put(54,57){\vector(2,-3){23}}
\put(22,40){${\delta {\Lambda}_{ij}}$}\put(68,40){${\delta g^{ij}}$}
\put(18,17){${S_{NG}}$}\put(78,17){${A^{{\rm I}}_{W}}$}
\put(20,10){\vector(0,1){5}}\put(80,10){\vector(0,1){5}}
\put(20,10){\line(1,0){20}}\put(60,10){\line(1,0){20}}
\put(46,9){\bf{dual}}
\end{picture}
\caption{\small Schematic relation of the actions: The parent action
$A^{{\rm III}}_{parent}$ shows that $S_{NG}$ and $A^{{\rm I}}_{W}$ are
dual to each other with respect to the inverse of the induced metric.}
\end{center}
\end{minipage}
\hspace{10mm}
\begin{minipage}{.45\linewidth}
\begin{center}
\setlength{\unitlength}{.7mm}
\begin{picture}(100,60)(10,10)
\put(46,60){$A^{{\rm IV}}_{parent}$}
\put(46,57){\vector(-2,-3){23}}\put(54,57){\vector(2,-3){23}}
\put(22,40){${\delta {\Lambda}}$}\put(68,40){${\delta {\Psi}}+{\delta {\Phi}}$}
\put(18,17){${A^{{\rm I}}_{W}}$}\put(78,17){${A^{{\rm II}}_{W}}$}
\put(20,10){\vector(0,1){5}}\put(80,10){\vector(0,1){5}}
\put(20,10){\line(1,0){20}}\put(60,10){\line(1,0){20}}
\put(46,9){\bf{dual}}
\end{picture}
\caption{\small Schematic relation of the actions: The parent action
$A^{{\rm IV}}_{parent}$ shows that $A^{{\rm I}}_{W}$ and $A^{{\rm II}}_{W}$
are dual to each other with respect to the scalar field.}
\end{center}
\end{minipage}
\end{figure}

The final task of this section is to write down the remaining parent
actions that can be constructed from the action set, ($S_{NG}$, $A_{P}$,
$A^{{\rm I}}_{W}$, $A^{{\rm II}}_{W}$):
\begin{equation}
A^{{\rm V}}_{parent}=-\frac{T_p}{2}\int d^{p+1}{\xi}\left[\sqrt{-G}
\left(-{\Phi}^{(p+3)/2}G_{ij}h^{ij}+(p+3){\Phi}^{(p+1)/2}\right)
 +{\Lambda}^{ij}\left(G_{ij}-g_{ij}\right)\right],
\end{equation}
and
\begin{equation}
A^{{\rm VI}}_{parent}=-T_{p}\int d^{p+1}{\xi}\left[\sqrt{-G}
 \left(\frac{1}{p+1}G_{ij}h^{ij}\right)^{-(p+1)/2}+{\Lambda}^{ij}
 \left(G_{ij}-g_{ij}\right)\right].
\end{equation}
Then we can calculate the corresponding dual actions. The calculation
procedure is nothing different from that described in sect.~4.
So we omit the details and just state the result that both $A^{{\rm I}}_{W}$
and $A^{{\rm II}}_{W}$ are dual to $S_{NG}$, but not to $A_{P}$, which
shows a kind of non-equality of $S_{NG}$ and $A_{P}$ in the set of dual
actions, ($S_{NG}$, $A_{P}$, $A^{{\rm I}}_{W}$, $A^{{\rm II}}_{W}$).
This non-equality is similar to that of $S_{NG}$ and $S_{P}$ described
in sect.~4. Furthermore, if we set $\Phi$ to a constant $b$ in
$A^{{\rm I}}_{W}$ in eq.~(\ref{sw1}), we obtain an infinite number of
$A_{P}$-type actions
\begin{equation}
A^{\prime}_{P}=-\frac{T^{{\prime}{\prime}}_p}{2}\int d^{p+1}{\xi}
\sqrt{-g}\left[-g_{ij}h^{ij}+(p+3)b^{-1}\right],
\end{equation}
where $T^{{\prime}{\prime}}_p=b^{(p+3)/2}T_p$, and the $b=1$ case
coincides with $A_{P}$. They are equivalent to the unique $A_{P}$ upon
using the equations of motion. The dualities between $A^{{\rm I}}_{W}$
and $S_{NG}$ and between $A^{{\rm II}}_{W}$ and $S_{NG}$ are illustrated
in Figs.~11 and 12, respectively.
\begin{figure}[htb]
\begin{minipage}{.45\linewidth}
\begin{center}
\setlength{\unitlength}{.7mm}
\begin{picture}(100,60)(10,10)
\put(46,60){$A^{{\rm V}}_{parent}$}
\put(46,57){\vector(-2,-3){23}}\put(54,57){\vector(2,-3){23}}
\put(22,40){${\delta {\Lambda}^{ij}}$}\put(68,40){${\delta G_{ij}}
+{\delta g_{ij}}$}
\put(18,17){${A^{{\rm I}}_{W}}$}\put(78,17){${S_{NG}}$}
\put(20,10){\vector(0,1){5}}\put(80,10){\vector(0,1){5}}
\put(20,10){\line(1,0){20}}\put(60,10){\line(1,0){20}}
\put(46,9){\bf{dual}}
\end{picture}
\caption{\small Schematic relation of the actions: The parent action
$A^{{\rm V}}_{parent}$ shows that $A^{{\rm I}}_{W}$ and $S_{NG}$ are
dual to each other with respect to the worldvolume metric.}
\end{center}
\end{minipage}
\hspace{10mm}
\begin{minipage}{.45\linewidth}
\begin{center}
\setlength{\unitlength}{.7mm}
\begin{picture}(100,60)(10,10)
\put(46,60){$A^{{\rm VI}}_{parent}$}
\put(46,57){\vector(-2,-3){23}}\put(54,57){\vector(2,-3){23}}
\put(22,40){${\delta {\Lambda}^{ij}}$}\put(68,40){${\delta G_{ij}}
+{\delta g_{ij}}$}
\put(18,17){${A^{{\rm II}}_{W}}$}\put(78,17){${S_{NG}}$}
\put(20,10){\vector(0,1){5}}\put(80,10){\vector(0,1){5}}
\put(20,10){\line(1,0){20}}\put(60,10){\line(1,0){20}}
\put(46,9){\bf{dual}}
\end{picture}
\caption{\small Schematic relation of the actions:
The parent action $A^{{\rm VI}}_{parent}$ shows that
$A^{{\rm II}}_{W}$ and $S_{NG}$ are dual to each other with respect to
the worldvolume metric.}
\end{center}
\end{minipage}
\end{figure}

\section{Conclusion}

By using the parent action approach of duality analysis, we construct
the parent actions that correspond to the four actions of bosonic $p$-branes
($p\geq 2$), $S_{NG}$, $S_{P}$, $S^{{\rm I}}_{W}$, and $S^{{\rm II}}_{W}$,
and establish the duality symmetries among them: (i) $S_{NG}$ and $S_P$ are
dual to each other (see Figs.~1 and 2),
(ii) the first and second Weyl-invariant actions with and without an
auxiliary scalar field, respectively, $S^{{\rm I}}_{W}$ and
$S^{{\rm II}}_{W}$, are also dual to each other (see
Figs.~3 and 4), and (iii) the two Weyl-invariant actions are dual to
$S_{NG}$, but not to $S_P$ (see Figs.~5 and 6).
This means that $S_{NG}$ and $S_P$, though dual to each
other, are not in an equal position even at the classical level in the set
of the four dual actions, ($S_{NG}$, $S_{P}$, $S^{{\rm I}}_{W}$,
$S^{{\rm II}}_{W}$). In fact, starting from the Weyl-invariant $p$-brane
action with an auxiliary scalar field, $S^{{\rm I}}_{W}$, we can deduce
an infinite number of so-called $S_P$-type actions~(\ref{pol}) that are
classically equivalent to the unique Nambu-Goto action upon using the
equations of motion. This also shows a kind of non-equality between $S_{NG}$
and $S_P$ from the point of view of the number of Lagrangians.

Furthermore, by proposing a parent action with the special
definition of the inverse of the induced metric, we discover a new bosonic
$p$-brane action (including the string case) which is nonpolynomial in
the derivatives of spacetime coordinates and equivalent to the Nambu-Goto
action classically. We subsequently obtain two corresponding Weyl-invariant
actions and establish the same duality symmetries as above (see Figs.~7 -- 12)
among the Nambu-Goto and three new actions, ($S_{NG}$, $A_{P}$,
$A^{{\rm I}}_{W}$, $A^{{\rm II}}_{W}$). The similar non-equality exists
between $S_{NG}$ and $A_{P}$ as that between $S_{NG}$ and $S_{P}$.
We may concisely illustrate the dualities established in the two sets of
dual actions, ($S_{NG}$, $S_{P}$, $S^{{\rm I}}_{W}$, $S^{{\rm II}}_{W}$)
and ($S_{NG}$, $A_{P}$, $A^{{\rm I}}_{W}$, $A^{{\rm II}}_{W}$) in
Figs.~13 and 14, respectively.
\begin{figure}[htb]
\begin{minipage}{.45\linewidth}
\begin{center}
\setlength{\unitlength}{.7mm}
\begin{picture}(50,60)(10,10)
\put(0,10){$S^{{\rm I}}_{W}$}
\put(50,10){$S^{{\rm II}}_{W}$}
\put(10,11){\vector(1,0){36}}
\put(46,11){\vector(-1,0){36}}
\put(0,60){$S_{NG}$}
\put(50,60){$S_{P}$}
\put(17,61.5){\vector(1,0){31}}
\put(46,61.5){\vector(-1,0){31}}
\put(2.5,18){\vector(0,1){39}}
\put(2.5,56){\vector(0,-1){39}}
\put(6,58){\vector(1,-1){44}}
\put(50,14){\vector(-1,1){44}}
\end{picture}
\caption{\small Dualities in the set of dual actions,
($S_{NG}$, $S_{P}$, $S^{{\rm I}}_{W}$, $S^{{\rm II}}_{W}$). No {\em direct}
dualities exist between $S^{{\rm I}}_{W}$ and $S_{P}$ and between
$S^{{\rm II}}_{W}$ and $S_{P}$.}
\end{center}
\end{minipage}
\hspace{10mm}
\begin{minipage}{.45\linewidth}
\begin{center}
\setlength{\unitlength}{.7mm}
\begin{picture}(50,60)(10,10)
\put(0,10){$A^{{\rm I}}_{W}$}
\put(50,10){$A^{{\rm II}}_{W}$}
\put(10,11){\vector(1,0){36}}
\put(46,11){\vector(-1,0){36}}
\put(0,60){$S_{NG}$}
\put(50,60){$A_{P}$}
\put(17,61.5){\vector(1,0){31}}
\put(46,61.5){\vector(-1,0){31}}
\put(2.5,18){\vector(0,1){39}}
\put(2.5,56){\vector(0,-1){39}}
\put(6,58){\vector(1,-1){44}}
\put(50,14){\vector(-1,1){44}}
\end{picture}
\caption{\small Dualities in the set of dual actions,
($S_{NG}$, $A_{P}$, $A^{{\rm I}}_{W}$, $A^{{\rm II}}_{W}$). No {\em direct}
dualities exist between $A^{{\rm I}}_{W}$ and $A_{P}$ and between
$A^{{\rm II}}_{W}$ and $A_{P}$.}
\end{center}
\end{minipage}
\end{figure}

Our result shows that the Nambu-Goto action is a kind of source from which
we can systematically derive its six dual actions in terms of the parent
action approach of duality analysis. In string theories, it was quite
useful to have various formulations of the actions for understanding the
geometrical meaning and covariant quantization. We hope that our systematic
investigation of the various formulations may be useful for these purposes.
Further studies of the new $p$-brane
actions are now being considered, such as the Hamiltonian analysis,
the supersymmetric extension, etc.
Finally we point out that our discussions can also be applied to
$Dp$-branes and similar conclusions can be made.

\vspace{10mm}
\noindent
{\bf Acknowledgments}
\par
We would like to thank S. Yahikozawa for valuable discussions on
Weyl-invariant $p$-brane actions.
This work was supported in part by Grants-in-Aid for Scientific Research
Nos.12640270 and 02041. Y.-G. M also acknowledges the support from
the National Natural Science Foundation of China under grant No.10275052.

\end{document}